\documentclass[12pt]{article}

\usepackage[totalheight = 23cm, totalwidth = 17cm]{geometry}
\usepackage{amssymb,amsmath,amsfonts,amsbsy,graphicx,bm}

\usepackage{color,cancel,ulem}

\newcommand{\dis}[1]{\begin{equation}\begin{split}#1\end{split}\end{equation}}

\begin{document}

\begin{titlepage}

\begin{center}

{\LARGE \bf 
Asymptotic bound on slow-roll parameter in stringy quintessence model
}

\vskip 1.0cm

{\large
Min-Seok Seo$^{a}$ 
}

\vskip 0.5cm

{\it
$^{a}$Department of Physics Education, Korea National University of Education,
\\ 
Cheongju 28173, Republic of Korea
}

\vskip 1.2cm

\end{center}

\begin{abstract}

 We study the late time behavior of the scalar part of the volume modulus and the dilaton in stringy quintessence model, focusing on their contributions   to the Hubble slow-roll parameter $\epsilon$ which  directly measures the deviation of the spacetime geometry from de Sitter   space. 
 When only one of the moduli is allowed to move, $\epsilon$ converges to the stable fixed point  at late time.
 The fixed point value is larger than $1$, thus the slow-roll cannot be realized.
 Moreover, if the decay rate of the quintessence potential is larger than some critical value, the positivity of the potential imposes that the stable fixed point value is just given by $3$, independent of the details of the moduli dynamics.
 Otherwise, the fixed point value coincides with the potential slow-roll parameter.
 When both  the volume modulus and the dilaton roll down the potential simultaneously, we can find the relation between the contributions of two moduli to $\epsilon$ satisfied at the fixed point. 
 In this case, the fixed point value is not in general  the simple sum of fixed point values in the single field case and cannot be larger than $3$.

\end{abstract}

\end{titlepage}

\newpage

\section{Introduction}

Construction of the model for the  universe compatible with the cosmological observations \cite{Planck:2018vyg} has been one of challenges in string phenomenology.
In particular, whereas models like the KKLT \cite{Kachru:2003aw} or the large volume scenario \cite{Balasubramanian:2005zx} have been proposed to realize the metastable de Sitter (dS) vacuum  which well describes the observed almost constant vacuum energy density, the suspicion has been raised that some unknown corrections may invalidate these models. 
It comes from the fact that in string theory, the parametric control is achieved in the asymptotic limits of the moduli space where the potential is dominated by a few runaway terms, but  the metastable dS vacuum requires that these terms are significantly corrected  \cite{Dine:1985he}.
Motivated by this, the `dS swampland conjecture' was proposed, which states that string theory does not admit dS vacua in any parametrically controlled regime of the moduli space \cite{Danielsson:2018ztv, Obied:2018sgi} (see,   \cite{Andriot:2018wzk, Garg:2018reu, Ooguri:2018wrx, Hebecker:2018vxz, Andriot:2018mav} for the refinement). 

If the conjecture is true, the accelerated expansion of the universe driven by an almost constant vacuum energy density may be explained by so-called the `quintessence model',  where the scalar field slowly  rolls down the runaway potential \cite{Peebles:1987ek, Ratra:1987rm, Caldwell:1997ii}. 
 This possibility stimulated extensive studies on the quintessence in the context of the string model building   \cite{Agrawal:2018own, Cicoli:2018kdo, DavidMarsh:2018etu, Hebecker:2019csg} (see also \cite{Hellerman:2001yi, Fischler:2001yj, Kaloper:2008qs, Cicoli:2012tz} for earlier discussions), in particular focusing on the behavior of the scalar fields in the   asymptotic region of the moduli space \cite{Olguin-Trejo:2018zun, ValeixoBento:2020ujr, Cicoli:2021fsd, Cicoli:2021skd, Brinkmann:2022oxy, Conlon:2022pnx, Rudelius:2022gbz, Calderon-Infante:2022nxb, Apers:2022cyl, Shiu:2023nph, Shiu:2023fhb, Cremonini:2023suw, Hebecker:2023qke,  VanRiet:2023cca, Revello:2023hro, Apers:2024ffe}.
 The remarkable claim made in recent works is that stringy quintessence models in the asymptotic region suffer from the no-go theorem : there is a lower bound of order one on the potential slow-roll parameter  which measures the slope of the potential in units of the Hubble scale.
 This is linked to the fact that for the string length and loop corrections to be suppressed, both the scalar part of the volume modulus (denoted by $\sigma$) and the dilaton (denoted by $s$) take the large values.
 In this case, the K\"ahler potential strongly restricts the slope of the potential in the direction of $\sigma$ and $s$, each of which is  too steep to realize the slow-roll   (see also \cite{Hertzberg:2007wc, Haque:2008jz, Flauger:2008ad, Caviezel:2009tu, Wrase:2010ew} for earlier discussions based on Type II string theory).
 In particular, it turns out that the leading no-scale structure of the potential is salient to obtain an order one  contribution of $\sigma$ to the potential slow-roll parameter.

 On the other hand, the time variation of the vacuum energy density can be more directly measured by the Hubble slow-roll parameter  defined by the  rate of change of the Hubble parameter $H$, 
 \dis{\epsilon=-\frac{\dot H}{H^2}.}
 Since $1/H$ is interpreted as the horizon radius, it is clear that $\epsilon$ measures the deviation of the spacetime geometry from dS space.
 In this regard, it can be claimed that $\epsilon$ rather than the potential slow-roll parameter is more appropriate parameter to describe the dS swampland conjecture, which essentially deals with the instability of the dS geometry (see \cite{Shiu:2023nph, Shiu:2023fhb} for earlier discussion).    
  \footnote{We note that in some models like the DBI  inflationary mechanism,  the effects of the non-negligible   higher-derivative terms can allow the sizeable potential slow-roll parameter while the value of  $\epsilon$ is kept smaller than $1$    \cite{Seo:2018abc, Mizuno:2019pcm}.}
   In the slow-roll approximation where the condition 
   \dis{1\gg \frac{\dot u}{H u}\gg \frac{\ddot u}{H^2 u}} 
 is satisfied for any scalar field $u$, $\epsilon$ can be well approximated by the potential slow-roll parameter.
 However, the bound on the potential slow-roll parameter claimed by the no-go theorem is of order one, which indicates that the slow-roll approximation does not hold, thus the potential slow-roll parameter is not necessarily identified with $\epsilon$.
 Then it is natural to ask if there is a bound on $\epsilon$ provided by the dynamics of $\sigma$ and $s$ in the asymptotic region, and if so, what is the value of the bound. 
 In this article, we address these questions by considering the specific model, the compactification of Type IIB string theory.
 
 In fact, the bound on $\epsilon$ in the generic quintessence model where a number of scalar fields roll down their own  runaway potentials was investigated in  \cite{Shiu:2023nph, Shiu:2023fhb}.
 They showed that $\epsilon$ is bounded from above as well as below,  and these bounds are determined by the positivity of the potential and the structure of the equations of motion in the asymptotic region.
 Our study corresponds to somewhat special case in which $\sigma$ and $s$ universally couple to the potential, i.e., every F-term potential term depends on these moduli in the same way, thus we cannot neglect them no matter how long time has passed.
 Then   the bound on $\epsilon$ contributed from $\sigma$ and $s$ can be interpreted  as the stable fixed point value. 
 Since we are familiar with the concept of the stable fixed point in the renormalization group analysis, our study is  expected to provide an intuitive way to interpret the results of \cite{Shiu:2023nph, Shiu:2023fhb}.

To proceed, we use the fact that when the quantum corrections  such as the non-perturbative effects  are negligible and the fluxes are turned off, the superpotential is independent of both $\sigma$ and $s$. 
  Then the F-term potential $V$ satisfies $u_i \partial_{u_i}V=-\beta_i V$  ($u_i=\sigma, s$) for some positive number $\beta_i$, which indeed is a typical feature of the runaway potential in the quintessence model.   
  If (for some reason) the sizeable but controllable quantum corrections are allowed or fluxes are turned on,  one of $\sigma$ and $s$ can be stabilized, while   another still rolls down the potential.
  \footnote{Of course, if both $\sigma$ and $s$ are stabilized, they never contribute to $\epsilon$.
  Moreover, there is no contribution of $\sigma$ to $\epsilon$ in the non-geometric compactifications that have no K\"ahler moduli  \cite{Calderon-Infante:2022nxb, Cremonini:2023suw}. }
 We first consider this simple case in Sec. \ref{Sec:Single}, based on the generic properties of the single field quintessence model summarized in Sec. \ref{Sec:GenQuint}.
 As we will see,  the value of  $\epsilon$ in this case converges to some stable fixed point  at late time.
 Comparing with the bound on the potential slow-roll parameter in the no-go theorem, the fixed point value is similar  in size, but for $\sigma$, quite different in nature.
 More concretely, when we redefine the field for the canonical kinetic term, the potential decreases exponentially with respect to the redefined field.
 If the decay rate   of this potential is larger than some value, the positivity of the potential imposes that the fixed point value is fixed to $3$, independent of the decay rate.
 In our model, $\sigma$ corresponds to this case.
 This implies that unlike the bound on the potential slow-roll parameter in the no-go theorem, the fixed point value $3$ is not the result of  the no-scale structure.
 Meanwhile, when any quantum corrections are negligibly small and the fluxes are turned off, both $\sigma$ and $s$ are allowed to roll down the potential, which is visited in  Sec. \ref{Sec:multi}.  
 Even in this case, by the positivity of the potential, the fixed point value of $\epsilon$ cannot be larger than $3$ and it is in fact the stable fixed point value.   
 
 We close the introduction with the discussion on the phenomenological issues raised by using $\sigma$ or $s$ as the quintessence.
For this purpose, we note that the  4-dimensional (reduced) Planck scale $M_{\rm Pl}$ is given by 
\dis{M_{\rm Pl}^2=\frac{{\cal V}_0 \ell_s^6}{\kappa_{10}^2},}
where $\ell_s=m_s^{-1}$ is the string length scale, ${\cal V}_0\ell_s^6$ is the size of the internal volume (${\cal V}_0=\langle \sigma^{3/2}\rangle$), and $\kappa_{10}^2=g_s^2\ell_s^8/(4\pi)$ is the gravitational coupling in 10-dimensional supergravity ($g_s=\langle s^{-1}\rangle$ : the string coupling constant).
This also reads
\dis{m_s=\frac{g_s}{\sqrt{4\pi{\cal V}_0}}M_{\rm Pl}.}
When we compare physical parameters in different string vacua, it is typical to fix $M_{\rm Pl}$ to the observed value $2.4\times 10^{18}$GeV : for example, the distance conjecture claims that  in the vacuum corresponding to the limit $s\to \infty$ ($\sigma\to\infty$),  $m_s$ ($m_s$ as well as the Kaluza-Klein mass scale) giving the fixed value of $M_{\rm Pl}$ is extremely low \cite{Ooguri:2006in}.
In contrast, in our case, we are interested in the time evolution of $\sigma$ and $s$,  then it is reasonable to regard $m_s$ as the fixed fundamental scale and observe the time evolution of $M_{\rm Pl}$ (or equivalently, Newton's constant) resulting from the continuous changes in  $\sigma$ or $s$.
However, the time variation of $M_{\rm Pl}$ is strongly restricted to be $|\frac{1}{M_{\rm Pl}}\frac{d M_{\rm Pl}}{dt}|\lesssim 10^{-12}$ \cite{Uzan:2002vq, Uzan:2010pm}, which is not consistent with the non-negligible time variation of $\sigma$ or $s$.
 In addition, from the Einstein frame action, one finds that both $\sigma$ and $s$ universally couple to fields with Planckian strength.
 While this interaction can be regarded as the `fifth force', it also has not been found yet : scalar with mass less than meV must couple to matter much weaker than Planckian strength \cite{Adelberger:2003zx}.
These problems suggest that we need some other mechanisms which restrict the variation of $\sigma$ or $s$ to be much smaller than $M_{\rm Pl}$ (for the almost constant $M_{\rm Pl}$), or screen the fifth force.
For instance, it was found in \cite{Andriot:2024jsh} that when the radiation as well as the matter is taken into account,  the value of modulus is fixed to the almost constant value during the radiation/matter domination era due to the Hubble friction.
Moreover, the fifth force may be screened by the interaction between the moduli and the axion \cite{Brax:2023qyp}.
While we will not address these models in more detail as it is beyond the scope of this article, we note that the phenomenological issues above need to be taken into account for more reasonable model construction.

 \section{Fixed point in single field quintessence model}
 \label{Sec:GenQuint}
 
 We first consider the simplest quintessence model in which the single scalar field $u$ rolls down the positive runaway potential.
 Assuming the spatial homogeneity and isotropy, the metric can be  written as
 \dis{ds^2=-dt^2+a(t)^2 \delta_{ij}dx^i dx^j}
and $u$ depends only on $t$, then the action is given in the form of 
  \dis{S_{\rm quint}=\int d^4 x a^3\Big(\frac{M_{\rm Pl}^2}{2\alpha^2}\frac{\dot{u}^2}{u^2}-\frac{V_0}{u^\beta}\Big).\label{eq:qunitaction}}
We note that  in terms of the canonically normalized field  $\varphi=\frac{M_{\rm Pl}}{\alpha}\log u$ the action   can be rewritten as 
\dis{S_{\rm quint}=\int d^4 x a^3\Big(\frac12 {\dot \varphi}^2-V_0 e^{-\alpha\beta \frac{\varphi}{M_{\rm pl}}}\Big),\label{eq:canon}}
which shows that the potential decreases exponentially with respect to $\varphi$ with the decay rate  given by $\alpha\beta$.
In the following, our discussion will be made in terms of $u$ since \eqref{eq:qunitaction} is a typical form of the action for (the scalar part of) the modulus in the effective supergravity description of string theory.

 Taking the Einstein-Hilbert action into account in addition, we obtain following equations of motion :
  \dis{& 3M_{\rm Pl}^2H^2=\frac{M_{\rm Pl}^2}{2\alpha^2}\frac{\dot{ u}^2}{u^2} + \frac{V_0}{u^{\beta}},
  \\
  & 3M_{\rm Pl}^2H^2+2 M_{\rm Pl}^2 \dot{H}=-\frac{M_{\rm Pl}^2}{2\alpha^2}\frac{\dot{u}^2}{u^2} + \frac{V_0}{u^{\beta}},
  \\
  &\ddot{u}+3H\dot{u}-\frac{\dot{u}^2}{u }-\alpha^2\beta \frac{uV_0/u^\beta}{M_{\rm Pl}^2 }=0,}
  where $H=\dot{a}/a$ is the Hubble parameter.
 From the difference between the first two equations, one finds that 
  \dis{\epsilon=-\frac{\dot H}{H^2}=\frac{1}{2H^2M_{\rm Pl}^2}\frac{M_{\rm Pl}^2}{ \alpha^2}\frac{\dot{u}^2}{u^2}=\frac{1}{2\alpha^2}\Big(\frac{\dot u}{H u}\Big)^2.}
 Meanwhile, the first and the third equations can be rewritten as
  \dis{&1=\frac{1}{6\alpha^2}\Big(\frac{\dot u}{H u}\Big)^2+\frac{V_0/u^{\beta}}{3H^2 M_{\rm Pl}^2},
  \\
 &\frac{\ddot u}{H^2 u}+3 \frac{\dot u}{H u} -\Big(\frac{\dot u}{H u}\Big)^2 -\alpha^2\beta \frac{V_0/u^{\beta}}{H^2M_{\rm Pl}^2}=0, \label{eq:V&dds}}  
respectively,  which shows that $\frac{\ddot{u}}{H^2u}$ and $V=\frac{V_0}{u^\beta}$ can be written in terms of $\sqrt{\epsilon}=\frac{1}{\sqrt2\alpha}\frac{\dot u}{Hu}$
\footnote{While ${\dot u}$ may be negative, we concentrate on the positive ${\dot u}$ to discuss the case in which the moduli roll down the potential, i.e., the values of the moduli get larger as time goes on.} :
  \dis{&\frac{V}{3H^2M_{\rm Pl}^2}=\frac{V_0/u^{\beta}}{3H^2 M_{\rm Pl}^2}=1-\frac{\epsilon}{3},
  \\
  &\frac{\ddot{u}}{H^2u}=3\alpha^2\beta-3\sqrt2 \alpha \sqrt{\epsilon}+2\alpha^2\Big(1-\frac{\beta}{2}\Big)\epsilon.\label{eq:Vddotsigma}}
  Since $V\geq 0$, the first equation gives the bound $\epsilon \leq 3$.
The upper bound $\epsilon=3$ is saturated when $\frac{V}{3M_{\rm Pl}^2 H^2}=0$.
 If $u$ keeps rolling down the potential, one may na\"ively expect that at late time ($t\to\infty$), $V\to 0$, thus $\epsilon$ converges to $3$.
 However, as we will see, this is not always the case.

 \begin{figure}[!t]
 \begin{center}
 \includegraphics[width=0.45\textwidth]{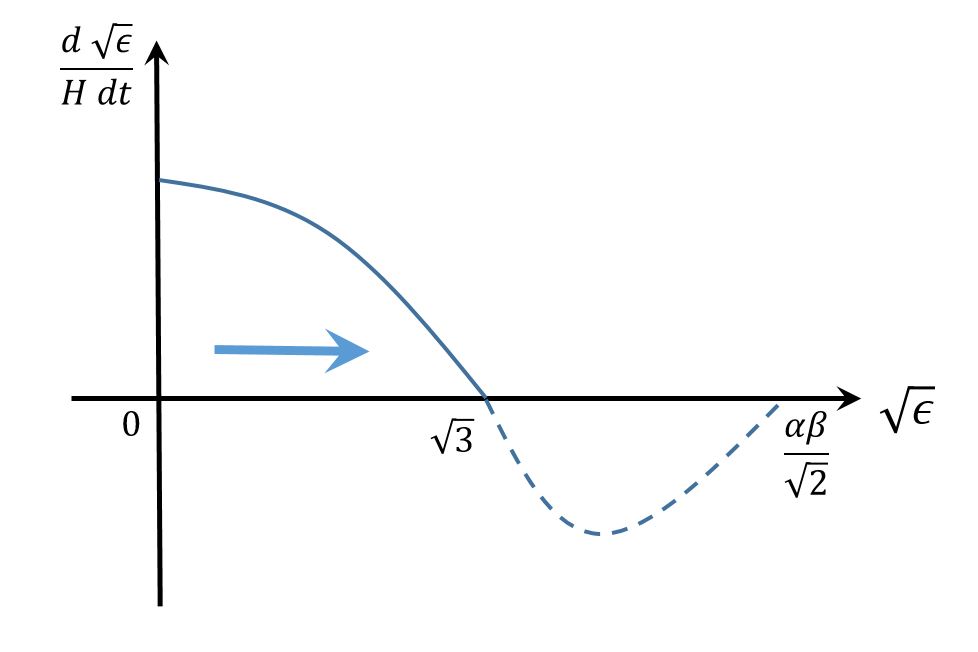}
 \includegraphics[width=0.45\textwidth]{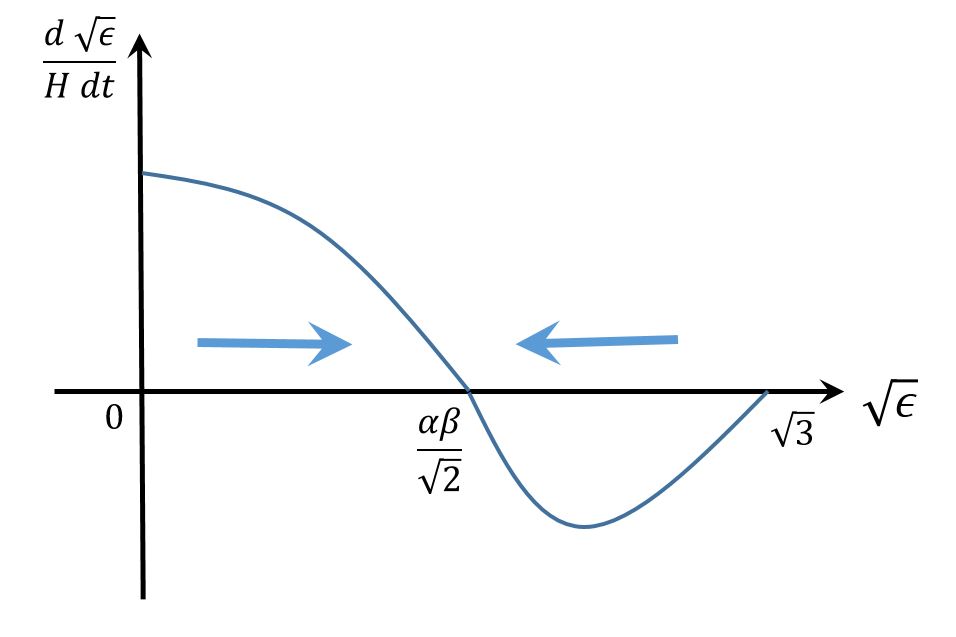}
 \end{center}
\caption{The variation of the rate $\frac{d}{Hdt}\sqrt{\epsilon}$ as a function of $\sqrt{\epsilon}$.
Left : For $\alpha\beta >\sqrt6$,  the positivity of the potential imposes  $\sqrt\epsilon<\sqrt3$, over which $\frac{d}{Hdt}\sqrt{\epsilon}$ is always positive.
Then regardless of the initial condition, $\sqrt\epsilon$ increases in time and eventually converges to $\sqrt3$.
Right : For   $\alpha\beta <\sqrt6$, $\frac{d}{Hdt}\sqrt{\epsilon}$ is negative (positive) for $\frac{\alpha\beta}{\sqrt2}<\sqrt\epsilon<\sqrt3$ ($0<\sqrt\epsilon<\frac{\alpha\beta}{\sqrt2}$)  in which case $\sqrt\epsilon$ decreases (increases) in time until it reaches the stable fixed point $\frac{\alpha\beta}{\sqrt2}$.
The direction of the time evolution of $\sqrt\epsilon$ is indicated by the arrow.
  }
\label{Fig:depsilon}
\end{figure}

 To see the time variation of $\epsilon$ in detail, consider the rate
 \dis{\frac{d}{Hdt}\sqrt{\epsilon}&=\frac{1}{\sqrt2 \alpha}\frac{d}{Hdt}\Big(\frac{\dot u}{H u}\Big)=\frac{1}{\sqrt2 \alpha}\Big[\frac{\ddot{u}}{H^2u}-\Big(\frac{\dot{u}}{H u}\Big)^2+\epsilon \frac{\dot{u}}{H u} \Big]
\\
&= (\sqrt\epsilon+\sqrt3)(\sqrt\epsilon-\sqrt3)\Big(\sqrt\epsilon-\frac{\alpha\beta}{\sqrt2}\Big),}
where   \eqref{eq:V&dds} is used for the last equality.
If $\alpha\beta >\sqrt6$, $\frac{d}{Hdt}\sqrt{\epsilon}$ is positive for $0\leq \sqrt\epsilon<\sqrt3$ and negative for  $\sqrt3<\sqrt\epsilon<\frac{\alpha\beta}{\sqrt2}$, but the latter region is not   physically meaningful due to the positivity of the potential $V=\frac{V_0}{u^\beta}$ : see the first equation in \eqref{eq:Vddotsigma}.
As depicted in the left panel of Fig. \ref{Fig:depsilon}, since $\frac{d}{Hdt}\sqrt{\epsilon}$ is always positive, regardless of the initial value, $\epsilon$ increases in time until it reaches $3$  at which the ratio $\frac{V}{3M_{\rm Pl}^2H^2}$  becomes $0$.
 Thus the upper bound on $\epsilon$ given by $3$ can be interpreted as a stable fixed point.
 We note that this stable fixed point value  does not depend on both $\alpha$ and $\beta$.
 This reflects the fact that for $\alpha\beta>\sqrt6$, the potential   quickly decreases to $0$ : as $u$ rolls down the potential, the ratio $\frac{V}{3M_{\rm Pl}^2H^2}$ converges to $0$ (where $\epsilon$ becomes $3$), while  both $V$ and $H$ decrease to $0$.
 In fact,   $\alpha\beta$ is nothing more than the decay rate of the potential with respect to  the canonically normalized field $\varphi=\frac{M_{\rm Pl}}{\alpha}\log u$ (see \eqref{eq:canon}).
 We also note that at $\epsilon=3$, $\frac{\ddot u}{H^2 u}=-(3\sqrt6-6\alpha)\alpha$ is   not necessarily suppressed compared to $\frac{\dot u}{H u}=\sqrt6 \alpha$ thus the slow-roll approximation is not guaranteed.  
 More precisely, $|\frac{\ddot u}{H^2u}|$ is larger than $\frac{\dot u}{Hu}$ for $\alpha <\sqrt{\frac23}$ and $\alpha >2\sqrt{\frac23}$.

 On the other hand, when $\alpha\beta=\sqrt6$,  the value of $\frac{\alpha\beta}{\sqrt2}$ at which $\frac{d}{Hdt}\sqrt\epsilon=0$ coincides with $\sqrt3$, thus $\frac{d}{Hdt}\sqrt\epsilon \geq 0$  for any   value of $\sqrt\epsilon$ in the physical region $0\leq \sqrt\epsilon \leq \sqrt3$ and $\frac{d}{Hdt}\sqrt\epsilon = 0$ only at $\epsilon=3$.
 Then the qualitative feature of the  time variation of $\epsilon$  is the same as that in the case of $\alpha\beta>\sqrt6$.
 \footnote{Comparing with  Sec. 4.3.1 of \cite{Bahamonde:2017ize}, this corresponds to the  fixed point $A_+$ at which the potential vanishes and the universe is dominated by the kinetic energy of the moduli.}

 Finally, when $\alpha\beta<\sqrt6$, $\frac{\alpha\beta}{\sqrt2}$   is smaller than $\sqrt3$ hence belongs to the physical region.
In this case, as can be found in the right panel in  Fig.  \ref{Fig:depsilon}, $\frac{d}{Hdt}\sqrt{\epsilon}$  is positive for $0<\sqrt\epsilon <\frac{\alpha\beta}{\sqrt2}$ and negative for $\frac{\alpha\beta}{\sqrt2}<\sqrt\epsilon <\sqrt3$. 
 Then $\epsilon=\frac{(\alpha\beta)^2}{2}$ corresponds to a stable fixed point : if the initial value of $\epsilon$ is smaller (larger) than $\frac{(\alpha\beta)^2}{2}$, $\epsilon$ increases (decreases) in time until it becomes $\frac{(\alpha\beta)^2}{2}$.  
Indeed, at the stable fixed point, the Hubble parameter,
\dis{H(t)= \frac{H(0)}{1+\frac{(\alpha\beta)^2}{2}H(0)t},}
behaves as $H(t)\sim \frac{2}{(\alpha\beta)^2}\frac{1}{t} \to 0$ at late time, and at the same time,  the potential
\dis{V=\Big(1-\frac{\epsilon}{3}\Big)3H^2M_{\rm Pl}^2 =\Big(1-\frac{(\alpha\beta)^2}{6}\Big)3M_{\rm Pl}^2\Big( \frac{H(0)}{1+\frac{(\alpha\beta)^2}{2}H(0)t}\Big)^2 }
also decreases to zero.
That is, while $\sigma$ rolls down the potential, both $V$ and  $H^2$  decrease to zero with the same rate   such that the ratio $\frac{V}{3M_{\rm Pl}^2H^2}$ is kept constant.
We also note that if $\alpha\beta\ll \sqrt2$ the stable fixed point value $\epsilon=\frac{(\alpha\beta)^2}{2}$ is much smaller than $1$.
Then the slow-roll condition is satisfied at late time provided  $\frac{\ddot\sigma}{H^2\sigma}=\alpha^4\beta^2\big(1-\frac{\beta}{2}\big)$ is much  smaller than $\frac{\dot\sigma}{H\sigma}=\alpha^2\beta$, in which case $V$ can be approximated by  $3 M_{\rm Pl}^2 H(t)^2$.

 We can compare $\epsilon$ with the `potential slow-roll parameter' given by
 \dis{\epsilon_V=\frac12 M_{\rm Pl}^2\Big(\frac{\alpha^2 u^2}{M_{\rm Pl}^2}\Big)\Big(\frac{dV/d u}{V}\Big)^2=\frac{(\alpha\beta)^2}{2},}
 where $\frac{\alpha^2 u^2}{M_{\rm Pl}^2}$ comes from the inverse of the   the K\"ahler metric. 
 \footnote{Obviously, the same value of $\epsilon_V$ can be obtained in terms of   the canonically normalized field $\varphi$ (see \eqref{eq:canon} for the action) by using the well known expression $\epsilon_V=\frac12 M_{\rm Pl}^2 \Big(\frac{dV/d\varphi}{V}\Big)^2$, which is consistent with the fact that $\alpha\beta$ is the decay rate of the potential.}
 Then one finds that if $\epsilon=\epsilon_V$, it is nothing more than the value of $\epsilon$ at which $\frac{d}{Hdt}\sqrt{\epsilon}$ becomes $0$.
 Whether it is a physical stable fixed point  depends on the size of $\alpha\beta$.
That is, while we can approximate  $\epsilon$ by $\epsilon_V$ when the slow-roll condition   is satisfied, the exact equality  $\epsilon_V= \epsilon$ is satisfied at the stable fixed point provided $\alpha\beta\leq \sqrt6$.
\footnote{See also \cite{Shiu:2023nph} which discussed this result in the context of the scaling solution, where the scale factor is of power-law form, i.e., $a(t)=a_0(t/t_0)^p$ for some positive power $p$. }

 \section{Volume modulus and dilaton runaways }
  \label{Sec:Single}

 We now move onto the contributions of (the scalar part of) the volume modulus and the  dilaton to $\epsilon$  in the asymptotic region when only one of them rolls down the potential.
 For this purpose,  we consider the compactification of Type IIB string theory with a single  K\"ahler modulus,   the volume modulus $\rho$.
 In this case, the K\"ahler potential  is given by
  \dis{K/M_{\rm Pl}^2=-3\log\big[-i(\rho-\overline{\rho})\big]-\log\big[-i(\tau-\overline{\tau})\big]+K_0/M_{\rm Pl}^2,}
  where $\tau$ is the axio-dilaton and $K_0$ depends on the complex structure moduli.
Meanwhile,  the superpotential $W$ is independent of $\rho$ in the absence of the quantum corrections such as the non-perturbative effects.
   Then the action for the moduli $\phi^I$ (which include $\rho$ and $\tau$, as well as the complex structure moduli)  is  
\dis{S[\phi^I]=\int d^4x a^3 \big( K_{I{\overline J}}{\dot \phi}^I {\dot {\overline\phi}}^{\overline J} -V(\phi^I)\big), }
where  the potential is given by
 \dis{V=e^{K/M_{\rm Pl}^2}\Big(K^{I{\overline{J}}} D_IWD_{\overline J} \overline{W}-\frac{3}{M_{\rm Pl}^2}|W|^2\Big),}
 with $D_I W=\partial_I W+\frac{1}{M_{\rm Pl}^2}\partial_I K$.
Denoting the scalar part of  $\rho$ and $\tau$ by $\sigma$ and $s$, respectively (that is, $\rho=i\sigma+\theta$ and $\tau=i s+a$), the volume of the internal manifold in string units is given by $\sigma^{3/2}$ and the string coupling constant $g_s$ is identified with $s^{-1}$.
Since $K^{\rho {\overline \rho}}K_\rho K_{\overline \rho}=3 M_{\rm Pl}^2$, in the absence of the quantum corrections,  the potential exhibits the no-scale structure,
 \dis{V=\frac{e^{K_0/M_{\rm Pl}^2}}{2s \times 8\sigma^3}K^{i\overline{j}}D_i W D_{\overline j}\overline{W},\label{eq:FinPot}}
 where $i, j$ run over moduli other than $\sigma$.

 \subsection{Volume modulus}
 
 We first assume that $s$ is stabilized, i.e., the value of $g_s$ is fixed, and investigate the rolling of $\sigma$.
 While the potential can be written as $V(\sigma)=V_0/\sigma^3$, for the comparison with the more suppressed potential (which violates the no-scale structure), we consider the potential given by $V=V_0/\sigma^{3+q}$, where $q\geq 0$. 
 Then the action for $\sigma$ is
  \dis{S_\sigma=\int d^4x a^3\Big(\frac34 M_{\rm Pl}^2\frac{\dot{ \sigma}^2}{\sigma^2}-\frac{V_0}{\sigma^{3+q}}\Big),}
which is the same form as \eqref{eq:qunitaction}  with $\alpha=\sqrt{\frac23}$ and $\beta=3+q$.
In other words, the field redefinition for the canonical kinetic term is given by $\varphi=\sqrt{\frac32}M_{\rm Pl}\log\sigma$.
The slow-roll parameter in this case is given by
  \dis{\epsilon=  \frac{2K_{\rho{\overline \rho}}\dot{\sigma^2}}{ 2H^2M_{\rm Pl}^2}=\frac34 \Big(\frac{\dot \sigma}{H\sigma}\Big)^2,}
  whereas the potential and $\frac{\ddot\sigma}{H^2\sigma}$ is written as  
  \dis{&\frac{V}{3H^2M_{\rm Pl}^2}=\frac{V_0/\sigma^{3+q}}{3H^2 M_{\rm Pl}^2}=1-\frac{\epsilon}{3},
  \\
  &\frac{\ddot{\sigma}}{H^2\sigma}=6\Big(1+\frac{q}{3}\Big)-2\sqrt3 \sqrt{\epsilon}-\frac23(1+q) \epsilon,\label{eq:Vddsig}}   
respectively.

The values of $\alpha$ and $\beta$ for $\sigma$ satisfy $\alpha\beta=\sqrt6\big(1+\frac{q}{3}\big)$, thus according to the discussion in Sec. \ref{Sec:GenQuint}, the rate of the change of $\sqrt\epsilon$ in time is given by
 \dis{\frac{d}{Hdt}\sqrt{\epsilon}&=\frac{\sqrt3}{2}\frac{d}{Hdt}\Big(\frac{\dot\sigma}{H\sigma}\Big)=(\sqrt{\epsilon}+\sqrt3)(\sqrt{\epsilon}-\sqrt3)\Big(\sqrt\epsilon-\sqrt3\Big(1+\frac{q}{3}\Big)\Big),}
and the positivity of $V$ imposes that the viable range of $\sqrt{\epsilon}$ is restricted to  $0 \leq \sqrt{\epsilon} \leq\sqrt3$.
Since   $\frac{d}{Hdt}\sqrt{\epsilon}$ is always positive in this region,  as  time goes on, $\epsilon$ increases toward the fixed point $\epsilon=3$, which is satisfied at $V=0$, or equivalently, $\sigma =\infty$.
Moreover, since   the fixed point  value   $\epsilon=3$  is larger than $1$  and the absolute value of $\frac{\ddot \sigma}{H^2\sigma}=-2$ is  the same  as that of $\frac{\dot \sigma}{H \sigma}=2$, the slow-roll condition is not satisfied.


The fact that  the fixed point value $\epsilon=3$ is independent of the choice of $q$ may be compared to the potential slow-roll parameter 
 \dis{\epsilon_V=\frac12 M_{\rm Pl}^2\frac{K^{\rho \overline{\rho}}}{2}\Big(\frac{dV/d\sigma}{V}\Big)^2=3\Big(1+\frac{q}{3}\Big)^2.}
When $q>0$, $\epsilon_V$ cannot be the viable value of  $\epsilon$ since $\epsilon=\epsilon_V$ gives $\frac{V}{3 H^2M_{\rm Pl}^2}=-\frac{q}{3}\big(2+\frac{q}{3}\big)<0$ (see the first equation in \eqref{eq:Vddsig}), which is not compatible with the positivity of the potential.
 Meanwhile, when $q=0$, $\epsilon_V$ coincides with the fixed point value $\epsilon=3$.
 We note that $\epsilon_V$ is given by $3$   only if the K\"ahler potential depends on $\sigma$ as $-3\log[2\sigma]$ (not, say, $-(3 +q)\log[2\sigma]$) and $W$ is independent of $\sigma$, or equivalently, the potential exhibits the no-scale structure thus $q$ is exactly $0$.
 In contrast, the fixed point value of $\epsilon$ is $3$ regardless of the value of $q$, and it indeed originates from the more generic condition, the positivity of the potential (given by the first equation in \eqref{eq:Vddotsigma} or \eqref{eq:Vddsig}), not the no-scale structure.

 \subsection{Dilaton}

 We now consider the case in which the value of $\sigma$ is fixed but $s$ rolls down the potential.
 When the fluxes are turned off, $W$ is independent of $\tau$ (hence $s$) as well, then from $K^{\tau {\overline \tau}}K_\tau K_{\overline \tau}= M_{\rm Pl}^2$ the potential can be written as $V=V_0/s$.
 \footnote{Of course, for $\sigma$ to be stabilized, the no-scale structure is violated by the non-perturbative effects or the supersymmetry breaking so \eqref{eq:FinPot} cannot be directly used.
 But here we assume that the potential term deviates from \eqref{eq:FinPot} is suppressed. }
 From this,  the action for $s$ is given in the form of \eqref{eq:qunitaction} with  $\alpha=\sqrt2$ and $\beta=1$,
   \dis{S_s=\int d^4x a^3\Big(\frac14 M_{\rm Pl}^2\frac{\dot{ s}^2}{s^2}-\frac{V_0}{s}\Big),}
 and the redefined field  for the canonical kinetic term is $\varphi=\sqrt{\frac12}M_{\rm Pl}\log s$.
Thus, the equations of motion give the relations
  \dis{&\epsilon=  \frac{2K_{\tau{\overline \tau}}\dot{s^2}}{ 2H^2M_{\rm Pl}^2}=\frac14 \Big(\frac{\dot s}{Hs}\Big)^2,
  \\
  &\frac{V}{3H^2M_{\rm Pl}^2}=\frac{V_0/s}{3H^2 M_{\rm Pl}^2}=1-\frac{\epsilon}{3},
  \\
  &\frac{\ddot{s}}{H^2s}=6 -6 \sqrt{\epsilon}+2\epsilon,}   
  respectively.
  Since $\alpha\beta=\sqrt2 <\sqrt3$, the rate of change of $\sqrt\epsilon$ in time is
  \dis{\frac{d}{Hdt}\sqrt\epsilon= \frac12\frac{d}{Hdt}\Big(\frac{\dot s}{Hs}\Big)=(\sqrt\epsilon+\sqrt3)(\sqrt\epsilon-1)(\sqrt\epsilon-\sqrt3).}
  This shows that $\epsilon=1$ corresponds to the stable fixed point : at late time $\epsilon$ converges to $1$ regardless of its  initial value in the physical region $0\leq \epsilon \leq 3$.
 From the equations of motion one finds that $\frac{V}{3 H^2M_{\rm Pl}^2}=\frac13$ and $\frac{\dot s}{Hs}=\frac{\ddot s}{H^2s}=2$ are satisfied at the fixed point, which indicates that the slow-roll approximation is not valid.
 We also note that the value of the potential slow-roll parameter
 \dis{\epsilon_V=\frac12 M_{\rm Pl}^2\frac{K^{\tau \overline{\tau}}}{2}\Big(\frac{dV/ds}{V}\Big)^2=1}
 coincides with $\epsilon$ at the fixed point.

 \section{Fixed point in multifield quintessence model}
 \label{Sec:multi}
 
 So far we have considered the simplest quintessence model in which only a single scalar field rolls down the runaway potential. 
 When more than two scalar fields $u_i$ ($i=1,\cdots, N$) simultaneously roll down the potential, we need to consider the generalized equations of motion, 
   \dis{& 3M_{\rm Pl}^2H^2=\sum_i\frac{M_{\rm Pl}^2}{2\alpha_i^2}\frac{\dot{u}_i^2}{u_i^2} + V,
  \\
  & 3M_{\rm Pl}^2H^2+2 M_{\rm Pl}^2 \dot{H}=-\sum_i\frac{M_{\rm Pl}^2}{2\alpha_i^2}\frac{\dot{u}_i^2}{u_i^2} + V,
  \\
  &\ddot{u}_i+3H\dot{u}_i-\frac{\dot{u}_i^2}{u_i }+\frac{\alpha_i^2}{M_{\rm Pl}^2}  u_i^2\frac{\partial V }{ \partial u_i}=0,}
  where we assume that the kinetic mixing between fields is absent.
  \footnote{In the presence of the kinetic mixing, the kinetic term can be written as $\frac{1}{2}G_{ij}\dot{u}^i\dot{u}^i$, which replaces  $\sum_i\frac{M_{\rm Pl}^2}{2\alpha_i^2}\frac{\dot{ u}_i^2}{u_i^2}$ in the first two equations.
  Then the third equation becomes
  \dis{{\ddot u}^i +\Gamma^{i}_{~jk}\dot{u}^j\dot{u}^k+3 H\dot{u}^i +G^{ij}\partial_j V=0,}
  where $\Gamma^{i}_{~jk}=\frac12G^{il}[\partial_j G_{lk}+\partial_k G_{lj}-\partial_l G_{jk}]$.
   }
  Then they lead to   
  \dis{&\epsilon=-\frac{\dot H}{H^2}=\sum_i \frac{1}{2\alpha_i^2}\Big(\frac{\dot{u}_i}{H u_i}\Big)^2 \equiv \sum_i\epsilon_i,
  \\
  &\frac{V}{3H^2M_{\rm Pl}^2}=1-\frac{\epsilon}{3},}
  where $\epsilon_i =\frac{1}{2\alpha_i^2}\big(\frac{\dot{u}_i}{Hu_i}\big)^2$.
 Moreover, when the potential satisfies $u_i\partial_i V=-\beta_i V$, i.e., exhibits the runaway behavior with respect to each of $u_i$, we obtain the relation
  \dis{
  &\frac{\ddot{u}_i}{H^2u_i}=3\alpha_i^2\beta_i-3\sqrt2 \alpha_i \sqrt{\epsilon_i}+2\alpha_i^2\epsilon_i- \alpha_i^2\beta_i\epsilon.}
 Since $\epsilon$, as well as $\epsilon_i$,  appears in the above equation (see the last term), the dynamics of any one of moduli   is affected by that of others.
 Of course, in the slow roll approximation in which  $\frac{\ddot{u}_i}{H^2u_i}$, $\epsilon$, and $\epsilon_i$ are suppressed compared to $\sqrt{\epsilon_i}$, each of $\epsilon_i$ is decoupled from the rest, giving $\epsilon_i\simeq \frac{(\alpha_i\beta_i)^2}{2}$ provided $\alpha_i\beta_i\ll 1$.
  
 \begin{figure}[!t]
 \begin{center}
 \includegraphics[width=0.45\textwidth]{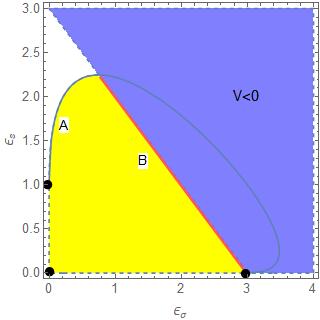}
 \end{center}
\caption{The behavior of the rate $\frac{d\epsilon}{Hdt}$ with respect to $\epsilon_\sigma$ and $\epsilon_s$.
The curve consists of points satisfying $\epsilon=\epsilon_\sigma+\epsilon_s=\sum_i\frac{\alpha_i\beta_i}{\sqrt2}\sqrt{\epsilon_i}$.
The blue region  corresponds to $\epsilon=\epsilon_\sigma+\epsilon_s>3$, which is excluded by the positivity of the potential. 
The yellow region satisfies both $\epsilon<3$ and $\epsilon<\sum_i\frac{\alpha_i\beta_i}{\sqrt2}\sqrt{\epsilon_i}$  (hence $\frac{d\epsilon}{Hdt}>0$).
 Then the part of the boundary of the yellow region satisfying $\epsilon=\sum_i\frac{\alpha_i\beta_i}{\sqrt2}\sqrt{\epsilon_i}$ or $\epsilon=3$ consists of  the stable fixed points.
  }
\label{Fig:multifield}
\end{figure}

 Meanwhile, from 
 \dis{\frac{d\epsilon}{Hdt}=\sum_i\frac{d\epsilon_i}{Hdt}&= \sum_i \frac{1}{\alpha_i^2}\Big(\frac{\dot{u}_i}{Hu_i}\Big)\Big(\frac{\ddot{u}_i}{H^2u_i}-\Big(\frac{\dot{u}_i}{Hu_i}\Big)^2+\epsilon\frac{\dot{u}_i}{Hu_i}\Big)
 \\
 &=2(\epsilon-3)\Big(\epsilon-\sum_i\frac{\alpha_i\beta_i}{\sqrt2}\sqrt{\epsilon_i}\Big),}
 we find that the value of $\epsilon$  at the fixed point $\frac{d\epsilon}{Hdt}=0$ is given by
 \dis{\epsilon={\rm min}\Big[3, \sum_i\frac{\alpha_i\beta_i}{\sqrt2}\sqrt{\epsilon_i}\Big].\label{eq:fixedmulti} } 
 In particular, a set of values $\{\epsilon_i =\frac{(\alpha_i\beta_i)^2}{2}\}$  can be the  fixed point so far as $\epsilon=\sum_i \frac{(\alpha_i\beta_i)^2}{2}$ is smaller than $3$ (otherwise, it contradicts to the positivity of the potential).
 In this case, each of $\epsilon_i$ coincides with the contribution of $u_i$ to the potential slow roll parameter,
 \dis{\epsilon_{V,i}=\frac12 M_{\rm Pl}^2\Big(\frac{\alpha_i^2 u_i^2}{M_{\rm Pl}^2}\Big)\Big(\frac{dV/d u_i}{V}\Big)^2=\frac{(\alpha_i \beta_i)^2}{2}.}
Moreover, unlike the single field case, the vanishing of $\frac{d\epsilon}{Hdt}$ does not guarantee that  
\dis{\frac{d}{H dt}\sqrt{\epsilon_i}=\sqrt2 \alpha_i\frac{d}{H dt}\Big(\frac{\dot{u}_i}{Hu_i}\Big)=2 \alpha_i^2 (\epsilon-3)\Big(\sqrt{\epsilon_i}-\frac{\alpha_i\beta_i}{\sqrt2}\Big)\label{eq:each}} 
must vanish for every moduli :  we just require that their combination  $\frac{d\epsilon}{Hdt}$ vanishes.
This means that a set of values $\{\epsilon_i\}$ satisfying $\frac{d\epsilon}{Hdt}=0$ in general evolves in time.
If this `point' $\{\epsilon_i\}$ no longer maintains the constant value of $\epsilon$ as a result of the time evolution, it cannot be the fixed point even if $\frac{d\epsilon}{Hdt}=0$ there.
Indeed, in contrast to the single field case, we can have another direction in the multi-dimensional field space which allows the point to evolve across the point satisfying $\frac{d\epsilon}{Hdt}=0$ (hence the sign of $\frac{d\epsilon}{Hdt}$ is changed), instead of going back to the point  satisfying $\frac{d\epsilon}{Hdt}=0$.
Moreover, even if the point satisfying $\frac{d\epsilon}{Hdt}=0$ is the fixed point, when the values of $\epsilon_i$ near the fixed point move away from the fixed point values through the time evolution, the fixed point in this case is not stable.
On the other hand, as evident from \eqref{eq:each}, when $\{\epsilon_i\}$ is on the surface $\epsilon=\sum_i\epsilon_i=3$, each of $\epsilon_i$ as well as  $\epsilon$ does not evolve in time, indicating that it is a fixed point.
Since the value of $\epsilon$ cannot be larger than $3$, it can be stable if it is on the boundary of the region  $\frac{d\epsilon}{Hdt}>0$.

 We now investigate the contributions of $\sigma$ and $s$ to $\epsilon$.
 In this case $(\alpha_i, \beta_i)$ ($i=\sigma, s$) are given by $(\sqrt{\frac23}, 3)$ for $\sigma$ and $(\sqrt{2}, 1)$ for $s$,   respectively.
 From this and $\epsilon=\epsilon_\sigma+\epsilon_s$ we obtain
 \dis{\frac{d\epsilon}{Hdt}&=2(\epsilon-3)\Big(\epsilon-(\sqrt3\sqrt{\epsilon_\sigma}+\sqrt{\epsilon_s})\Big)
 \\
 &=2(\epsilon-3)\big(\sqrt{\epsilon_\sigma}(\sqrt{\epsilon_\sigma}-\sqrt3)+\sqrt{\epsilon_s}(\sqrt{\epsilon_s}-1)\big).}
 Since the positivity of the potential imposes $\epsilon\leq 3$, $\frac{d\epsilon}{Hdt}$ is positive when both $\epsilon<3$ and $\epsilon<\sqrt3\sqrt{\epsilon_\sigma}+\sqrt{\epsilon_s}$ are satisfied, which corresponds to the yellow region in Fig. \ref{Fig:multifield}.
 Then we expect that the fixed point belongs to the boundary of this region, which consists of the curve A (satisfying $\epsilon=\sum_i\frac{\alpha_i\beta_i}{\sqrt2}\sqrt{\epsilon_i}$) and a part of the line B (satisfying $\epsilon=3$).
 While $\frac{d \epsilon}{Hdt}=0$ at $(\epsilon_\sigma, \epsilon_s)=(0,0)$ as well, it corresponds to the unstable fixed point : a set of values $(\epsilon_\sigma, \epsilon_s)$ slightly deviates from $(0,0)$ is in the region $\frac{d\epsilon}{Hdt}>0$, hence tends to move away from $(0,0)$ as time goes on.
 We note that two curves $\epsilon=3$ and $\epsilon=\sqrt3\sqrt{\epsilon_\sigma}+\sqrt{\epsilon_s}$ intersect at $(\epsilon_\sigma, \epsilon_s)=(\frac34, \frac94)$ and $(3,0)$, at which $V=0$.
 Meanwhile, the points on the curve A except for $(\epsilon_\sigma, \epsilon_s)=(0,1)$ are not the fixed points due to the behavior of  $(\epsilon_\sigma, \epsilon_s)$, as explained in the previous paragraph.
 To see this, we recall that
 \dis{&\frac{d \epsilon_\sigma}{Hdt}=2(3-\epsilon)\sqrt{\epsilon_\sigma}(\sqrt3-\sqrt{\epsilon_\sigma}),
 \\
 &\frac{d \epsilon_s}{Hdt}=2(3-\epsilon)\sqrt{\epsilon_s}(1-\sqrt{\epsilon_s}).}
 Since $3-\epsilon$ is positive, we expect that on the curve A where $\epsilon_\sigma<3$ and $\epsilon_s>1$ are satisfied, $\frac{d \epsilon_\sigma}{Hdt}>0$ and $\frac{d \epsilon_s}{Hdt}<0$.
 In other words, $\epsilon_\sigma$ ($\epsilon_s$) on the curve A increases (decreases) in time, hence $(\epsilon_\sigma, \epsilon_s)$ evolves into the region $\frac{d \epsilon}{Hdt}>0$, which cannot happen at the fixed point. 
 Moreover, even though $(\epsilon_\sigma, \epsilon_s)=(0, 1)$ is the fixed point, it is unstable.
 To see this, we parametrize the values of $\sqrt{\epsilon_i}$ at the point near to $(0, 1)$ by $\sqrt\epsilon_\sigma= \delta x$ and $\sqrt\epsilon_s= 1+\delta y$, respectively, which lead to  $\frac{d \epsilon_\sigma}{Hdt} \simeq 4\sqrt3 \delta x$ and $\frac{d \epsilon_s}{Hdt} \simeq -4  \delta y$.
 This shows that while $\epsilon_s$ gets closer to the fixed point value $1$, $\epsilon_\sigma$ moves away from the fixed point value $0$.
 Therefore, the stable fixed point when both $\sigma$ and $s$ roll down the potential is restricted to the line B, at which $\epsilon$ is given by $3$, the largest value allowed by the positivity of the potential.  
 We also note that the point $(\epsilon_\sigma, \epsilon_s)=(3,1)$ at which $\epsilon_i=\epsilon_{V,i}$ is satisfied is excluded by the positivity of the potential. 
 This also indicates that the fixed point value is not necessarily the simple sum of the values obtained in the single field case.

 \section{Conclusions}
\label{sec:conclusion}

 In this short note, we consider the stringy quintessence model and investigate  the late time contributions of $\sigma$ (the scalar part of the volume modulus) and $s$  (the dilaton)  to the slow-roll parameter $\epsilon=-\frac{\dot H}{H^2}$  which directly measures the deviation of the geometry from dS space. 
 We point out that as time goes on, each of these contributions converges to the stable fixed point, at which the slow-roll approximation does not hold.
 In particular, in the single field model, it turns out that when the decay rate of the potential is larger than some critical value, the positivity of the potential imposes that the fixed point value is given by $3$, independent of the details of the dynamics. 
  This is somewhat different feature from the potential slow-roll parameter, and $\sigma$ corresponds to this case. 
  When both $\sigma$ and $s$ simultaneously roll down the potential,  we can find  the curve containing the fixed points in the plane of $\epsilon_\sigma$ and $\epsilon_s$, and as in the single field case,  there exists a direction   in which the fixed point value of $\epsilon$ is just given by $3$, the maximum value of $\epsilon$.
  In fact, the points on the part of the line $\epsilon=3$ corresponding to the boundary of the region $\frac{d\epsilon}{Hdt}>0$ are   the stable fixed points.
  Moreover, unlike the potential slow-roll parameter, the fixed point value in the multifield  case is not necessarily given by the sum of fixed point values in the single field case.
  
  The no-go theorem we have investigated indicates that despite the compatibility with the observations, both the metastable dS vacuum and the quintessence suffer from the parametric control problem.
  It is also remarkable that in order to resolve this problem, we need to understand the dynamics of $\sigma$ and $s$ more clearly.
  Presumably, there may be an accidental fine-tuning which prevents the instability of $\sigma$ and $s$ (for the instability issue originating from the mixing between $\sigma$ and $s$, see, \cite{Choi:2004sx, Seo:2021kyi}), or unknown quantum gravity reason which recovers the parametric control.

%

%


\appendix



\renewcommand{\theequation}{\Alph{section}.\arabic{equation}}


\subsection*{Acknowledgements} 

MS is grateful to Flavio Tonioni for explaining that the conclusions of this paper and those in  \cite{Shiu:2023nph, Shiu:2023fhb} are essentially equivalent.


\end{document}